# Bilayer Quantum Hall States in an n-type Wide Tellurium Quantum Well

Chang Niu[1,2], Gang Qiu[1,2], Yixiu Wang[3], Mengwei Si[1,2], Wenzhuo Wu[3]

and Peide D. Ye[1,2,*]

[1]*School of Electrical and Computer Engineering, Purdue University, West Lafayette, IN 47907, United States*

[2]*Birck Nanotechnology Center, Purdue University, West Lafayette, IN 47907, United States*

[3]*School of Industrial Engineering, Purdue University, West Lafayette, IN 47907, United States*

* Address correspondence to: yep@purdue.edu (P.D.Y.)

ABSTRACT

Tellurium (Te) is a narrow bandgap semiconductor with a unique chiral crystal structure. The topological nature of electrons in the Te conduction band can be studied by realizing n-type doping using atomic layer deposition (ALD) technique on two-dimensional (2D) Te film. In this work, we fabricated and measured the double-gated n-type Te Hall–bar devices, which can operate as two separate or coupled electron layers controlled by the top gate and back gate. Profound Shubnikov-de Haas (SdH) oscillations are observed in both top and bottom electron layers. Landau level hybridization between two layers, compound and charge-transferable bilayer quantum Hall states at filling factor $\nu = 4, 6$ and 8 are analyzed. Our work opens the door for the study of Weyl physics in coupled bilayer systems of 2D materials.

Bilayer systems that are made by confining electrons in two thin layers, exhibit a variety of phenomena, including integer and fractional quantum Hall effect[1–8], Hall drag[9], Coulomb drag[10] and exciton condensation[11–13]. The structure-induced additional layer degree of freedom provides an ideal platform for the research on strongly correlated electrons and multicomponent physics[14]. Compared to traditional double quantum wells based on GaAs and $Al_xGa_{1-x}As$[15], bilayer systems in wide quantum wells possess similar physics[2,16].

A bilayer electron system is realized in the Tellurium (Te) wide quantum well structure with two different electrical field induced two-dimensional electron gases (2DEGs) at the top and bottom surface of the Te crystal as illustrated in Figure 1(a). The nearly intrinsic Te between two conducting layers creates a potential barrier for the electron tunneling. By controlling the top and back gate bias and the magnetic field, electron density, tunneling interaction, interlayer and intralayer Coulomb interaction can be tuned in a single device.

The crystal structure of right-handed trigonal Te is shown in Figure 1(a). Covalently bonded Te atoms form a chiral atomic chain with threefold screw symmetry along z axis. The helical chains are arranged in a trigonal lattice through van der Waals interaction. Te is a narrow band gap semiconductor. The conduction band minimum which contains twofold valley and twofold spin degeneracies is at the corner of the first Brillouin zone (H and H')[17]. The unique crystal structure with chiral screw symmetry and the strong spin-orbit coupling of Te[18] bring exotic physical properties, including the camelback band structure in Te valence band[19,20], radial spin texture[21–23] and the presence of a Weyl node near the edge of conduction band[21,24–26]. The non-trivial π Berry phase of Weyl fermions in the quantum oscillation sequence is reported in our previous work[27].

Te material usually manifests p-type behaviour. A p-type accumulation layer[28,29] at the crystal surface emerges since its chemical potential is aligned near the valence band.

Atomic layer deposition (ALD) treatment removes the native oxide on the top surface and the positive fixed charges in ALD oxide converts the underneath 2D Te from p-type to n-type[30]. Hence, we can study the topological properties of Te conduction band in a controlled fashion. The device structure of double-gated Te field-effect transistors (FETs) is shown in Figure 1(b). Te flakes grown by hydrothermal method[31] are transferred onto 90nm $SiO_2$/Si substrate. Titanium (50 nm) ensures the ohmic electrical contact to n-type Te. 90 nm of $SiO_2$ and 20 nm $Al_2O_3$, grown by ALD at 200 °C, are used as the back and top gate dielectrics respectively. The optical image in Figure 1(c) shows a standard double-gated Hall-bar Te FET with the crystal orientations (determined in previous work[31]) indicated by the black arrows. The thickness of Te flake is 20 nm in this study. Transverse ($R_{xy}$) and longitudinal ($R_{xx}$) resistance are measured at cryogenic temperature by inducing a current along the Te chain direction z.

In this paper, we systematically study the quantum transport of the bilayer electron system in an n-type 2D Te wide quantum well described above. Temperature and electron density dependence of quantum oscillations are measured in the magnetic field up to 31 T. Landau level hybridization between two layers and well-developed charge-transferable and compound states quantum Hall plateaus of filling factor $\nu$ = 4, 6 and 8 are observed.

RESULTS AND DISCUSSION

Top gate voltage $V_{tg}$ and back gate voltage $V_{bg}$ dependence of the Te channel resistance $R_{xx}$ at cryogenic temperatures without magnetic field shown in Figure 1(c) confirms the n-type behavior of the double-gated Te device. Profound Shubnikov-De Haas (SdH) oscillations are observed when the magnetic field is applied perpendicular to the sample surface, due to the high electron mobility of the 2D Te. By sweeping the back gate voltage $V_{bg}$ (thus the carrier density) and the magnetic field B, a color mapping of $R_{xx}$ can be measured, as shown in Figure 1(d). Different Landau levels controlled by back gate voltage $V_{bg}$ are identified. We attribute the complicated oscillation pattern to spin and valley splitting[21] of the Te conduction band within a single layer of 2DEG.

To increase the resolution and visibility of the oscillation features, we take the second derivative of $R_{xx}$ with respect to the magnetic field B. The mapping of $-\partial^2 R_{xx}/\partial B^2$ gives the same information with more clarity compared to the original data ($R_{xx}$), as shown in Figure S1. Figure 2 shows the temperature-dependent mapping of $-\partial^2 R_{xx}/\partial B^2$ from 40 mK to 8 K at $V_{tg} = 0$ V. Because of the ALD treatment at Te top surface, electrons in the top layer experience stronger surface scattering than bottom layer electrons. Therefore, the electron mobility of the top layer is relatively low. With the increase of temperature from 40 mK (Figure 2(a)) to 1 K (Figure 2(b)), the bottom layer electron mobility drops fast, and when it is comparable with the electron mobility of the top layer, another set of quantum oscillations from the top surface appears, indicated by the red eye guidelines in Figure 2(c). Landau levels from different layers cross and form local maximum resistance points which is the bright area in the color mapping. We notice the small change of the Landau level crossing pattern at higher filling factors. This can also be explained by the Landau level hybridization between top and bottom layer 2DEGs (see Supporting Information Figure S2).

The SdH oscillations from top layer 2DEG can be tuned by changing the top gate voltage $V_{tg}$. Top layer carrier density is calculated from $B_F$ (the SdH oscillation frequency in 1/B) through $n = 4eB_F/h$ (here $h$ is the Planck constant) with twofold spin and twofold valley degeneracies[32]. The frequency of top layer 2DEG SdH oscillations (red dash lines, Figure 3(a)) at $V_{tg} = 6$ V and $V_{bg} = 15$ V is 35 T, indicating a top layer carrier density of $n_t = 3.4\times10^{12}$ cm$^{-2}$. The system returns to single bottom quantum well dominant behavior with $V_{tg} = -6$ V (Figure 3(b)), due to the relatively low mobility and density of the top layer electrons. Figure 3(c) shows the top and back gate voltage dependence of the SdH oscillations at B = 9 T. Two sets of SdH oscillations controlled separately by $V_{tg}$ and $V_{bg}$ form a discrete spectrum with Landau level crossing at each lattice point, which is the signature of bilayer electron system transport[1,33]. The top gate voltage $V_{tg}$ does not influence the bottom layer quantum oscillations (blue dash lines) owing to the charge screening from top layer electrons. On the other hand, the top layer electron density can be modulated by the back gate voltage $V_{bg}$. As the back gate voltage $V_{bg}$ increases, the top layer electron density also increases, causing the Landau level shift in Figure 3(a) and (c). The magnetic field B dependence of bilayer quantum oscillations can be seen in Figure S3 in the Supporting Information.

To further investigate the top layer quantum oscillations in double-gated Te FETs, we measured the color mapping of $R_{xx}$ (Figure 4(a)) by sweeping top gate voltage $V_{tg}$ and magnetic field B. SdH oscillations from bottom and top layer 2DEGs are separated when taking the second derivative of $R_{xx}$ with respect to the magnetic field B ($-\partial^2 R_{xx}/\partial B^2$, Figure 4(b)) and the top gate voltage $V_{tg}$ ($-\partial^2 R_{xx}/\partial V_{tg}^2$, Figure 4(c)). Vertical bright lines in Figure 4(b) show the domination of the quantum oscillations originated from bottom layer at $V_{bg} = 15V$. By eliminating the bottom layer contribution, Landau levels controlled by top gate voltage are shown in Figure 4(c). The lowest top surface Landau level we identified is n = 2 at the magnetic field of 12 T. Two gray curves in Figure 4(b) and (c) are $R_{xx}$ at $V_{tg} = 0$ V and $-\partial^2 R_{xx}/\partial V_{tg}^2$ at $V_{tg} = 3.1V$ respectively. The same oscillation frequency between two curves indicates the carrier density is balanced at two surfaces. The Landau level splitting (black arrows in Figure 4(b)) is not clear in top layer SdH oscillation, because of the large Landau level broadening at the top surface. Complete measurement of the top layer Landau fan diagram at different back gate voltage $V_{bg}$ is shown in Supporting Information Figure S5. Landau level sequences in top and bottom layer 2DEG are identical.

A bilayer electron system provides an additional "which-layer" degree of freedom in the third direction and gives rise to various bilayer quantum Hall states. In analogue to the spin property of electrons, the SU(2) pseudospin (P = $P_x$, $P_y$, $P_z$) structure is used to describe the bilayer quantum Hall system[1]. Here the up-pseudospin ($P_z$ = 1/2) component is the top-layer component and the down-pseudospin ($P_z$ = -1/2) component is the bottom-layer component. The electron transfer between two layers is a rotation of the pseudospin. When the tunneling interaction between two layers is large, one Landau level splits into symmetric and antisymmetric states separated by the tunneling gap $\Delta_{SAS}$. Charge-transferable states are realized when each level is filled. If $\Delta_{SAS}$ = 0 (no electron tunneling), the pseudospin contains only z component. Compound states are observed at filling factor $\nu = \nu^t + \nu^b$, where $\nu^t$ ($\nu^b$) is the filling factor of top (bottom) layer quantum Hall states[34].

Both compound and charge-transferable bilayer quantum Hall states of double-gated 2D n-type Te FETs are observed at low Landau levels shown in Figure 5, when a higher magnetic field is applied. Top gate voltage $V_{tg}$ dependence of transverse ($R_{xy}$, Figure 5(a) upper figure) and longitudinal ($R_{xx}$, Figure 5(a) lower figure) resistance as a function of magnetic field B are measured at $V_{bg}$ = 10 V. At a low magnetic field (below the red dash line), the frequency of the SdH oscillations originated from bottom layer 2DEG remains the same ($B_F$ = 15 T) at different top gate voltages, indicating a fixed bottom layer electron density ($n_b$ = 1.5×10$^{12}$ cm$^{-2}$) less sensitive to $V_{tg}$. However, at a high magnetic field, well-developed quantum Hall plateaus of filling factor $\nu$ = 4, 6 and 8 are observed by increasing the $V_{tg}$ (thus top layer electron density). Therefore, we conclude that quantum Hall states at filling factor 6 and 8 are compound states provided by both top and bottom layer 2DEG. The slight shift of the SdH oscillation minima and the change of the oscillation amplitude at lower magnetic field B indicate that the top gate has a small influence on the bottom layer electrons. The total electron density of the system determined by Hall measurement (Figure S6) linearly changes from 2.3×10$^{12}$ cm$^{-2}$ ($V_{tg}$ = -3 V) to 4.8×10$^{12}$ cm$^{-2}$ ($V_{tg}$ = 5 V). By comparing the total electron density and bottom layer quantum oscillation density, a carrier density balanced condition ($V_{bg}$ = 10 V, $V_{tg}$ = -1 V) is found between two 2DEGs. Figure 5(b) shows a color mapping of $R_{xx}$ by changing $V_{bg}$ and $V_{tg}$ at a magnetic field of 31 T. The number in the color mapping represents the filling factor of the quantum Hall states. Carrier density balanced condition between two 2DEGs (black dash line) is obtained by connecting two carrier density balanced point (black crosses: $V_{bg}$ = 10 V, $V_{tg}$ = -1 V; $V_{bg}$ = 15 V, $V_{tg}$ = 3.1 V). We identify two compound states ($v^t$, $v^b$) at filling factor $v$ = 4 (2, 2) and 6 (3, 3) along the carrier density balanced line in Figure 5(b). Quantum Hall states are observed not only on discrete lattice point ($v^t$, $v^b$) but also on elongated areas[34]. Instead of two independent single-layer quantum Hall states, the electrons in different layers at charge-transferable states are strongly correlated. The continuous quantum Hall states of filling factor 4 (from $V_{tg}$ = -6 V to 0 V) in Figure 5(b) and (c) indicates the charge is transferable between top and bottom layer 2DEG. The wave functions of electrons in top and bottom layer overlap when the gate bias is relatively low, giving rise to a finite tunneling gap $\Delta_{SAS}$. The charge transferable states can be observed in higher filling factors $v$ = 6, 8, 10, and 12 at the magnetic field of 12 T (see Supporting Information Figure S7), indicating a complicated Landau level hybridization between top and bottom layer 2DEG. Because of the Weyl node at the edge of the conduction band[21], the quantum oscillations of the Weyl fermions show a topological non-trivial $\pi$ phase shift[27]. In Figure S8, we extracted the Landau fan diagram of the top and bottom layer 2DEG by assigning the integer Landau level index n to the minima in quantum oscillations. The 0.5 intercept indicates the presence of the Weyl fermions in both top and bottom layers. Further improved device fabrication and clearer quantum oscillation features would provide a new interesting platform to investigate the correlated bilayer Weyl fermions and explore new topological physics.

CONCLUSION

In conclusion, double-gated n-type 2D Te FETs were fabricated and measured. SdH oscillations from top and bottom layer 2DEG controlled by the top and back gates are observed and separated. With the presence of an additional "which-layer" degree of freedom, the Landau level hybridization between top and bottom layer 2DEGs is studied. Compound and charge-transferable bilayer quantum Hall states of filling factor $v = 4, 6$ and 8 are identified in a wide Te quantum well structure at a high magnetic field. The realization of two well-controlled high mobility 2DEGs in a n-type wide Te quantum well provides an ideal platform for the study of the Weyl physics in bilayer systems.

REFERENCES

(1) Ezawa, Z. F. Quantum Hall Effects: Recent theoretical and experimental developments. *World Scientific Publishing Company* **2013**.

(2) Suen, Y. W.; Santos, M. B.; Shayegan, M. Correlated States of an Electron System in a Wide Quantum Well. *Phys. Rev. Lett.* **1992**, *69* (24), 3551–3554.

(3) Li, J. I. A.; Shi, Q.; Zeng, Y.; Watanabe, K.; Taniguchi, T.; Hone, J.; Dean, C. R. Pairing States of Composite Fermions in Double-Layer Graphene. *Nat. Phys.* **2019**, *15* (9), 898–903.

(4) Suen, Y. W.; Engel, L. W.; Santos, M. B.; Shayegan, M.; Tsui, D. C. Observation of a =1/2 Fractional Quantum Hall State in a Double-Layer Electron System. *Phys. Rev. Lett.* **1992**, *68* (9), 1379–1382.

(5) Murphy, S. Q.; Eisenstein, J. P.; Boebinger, G. S.; Pfeiffer, L. N.; West, K. W. Many-Body Integer Quantum Hall Effect: Evidence for New Phase Transitions. *Phys. Rev. Lett.* **1994**, *72* (5), 728–731.

(6) Sawada, A.; Ezawa, Z. F.; Ohno, H.; Horikoshi, Y.; Ohno, Y.; Kishimoto, S.;

Matsukura, F.; Yasumoto, M.; Urayama, A. Phase Transition in the N=2 Bilayer Quantum Hall State. *Phys. Rev. Lett.* **1998**, *80* (20), 4534–4537.

(7) Eisenstein, J. P.; Boebinger, G. S.; Pfeiffer, L. N.; West, K. W.; He, S. New Fractional Quantum Hall State in Double-Layer Two-Dimensional Electron Systems. *Phys. Rev. Lett.* **1992**, *68* (9), 1383–1386.

(8) Spielman, I. B.; Eisenstein, J. P.; Pfeiffer, L. N.; West, K. W. Resonantly Enhanced Tunneling in a Double Layer Quantum Hall Ferromagnet. *Phys. Rev. Lett.* **2000**, *84* (25), 5808–5811.

(9) Liu, X.; Watanabe, K.; Taniguchi, T.; Halperin, B. I.; Kim, P. Quantum Hall Drag of Exciton Condensate in Graphene. *Nat. Phys.* **2017**, *13* (8), 746–750.

(10) Kim, S.; Jo, I.; Nah, J.; Yao, Z.; Banerjee, S. K.; Tutuc, E. Coulomb Drag of Massless Fermions in Graphene. *Phys. Rev. B* **2011**, *83* (16), 161401.

(11) Ezawa, Z. F.; Iwazaki, A. Quantum Hall Liquid, Josephson Effect, and Hierarchy in a Double-Layer Electron System. *Phys. Rev. B* **1993**, *47* (12), 7295–7311.

(12) Li, J. I. A.; Taniguchi, T.; Watanabe, K.; Hone, J.; Dean, C. R. Excitonic Superfluid Phase in Double Bilayer Graphene. *Nat. Phys.* **2017**, *13* (8), 751–755.

(13) Eisenstein, J. P.; MacDonald, A. H. Bose-Einstein Condensation of Excitons in Bilayer Electron Systems. *Nature* **2004**, *432*, 691–694.

(14) Rasolt, M.; MacDonald, A. H. Collective Excitations in the Fractional Quantum Hall Effect of a Multicomponent Fermion System. *Phys. Rev. B* **1986**, *34* (8), 5530–5539.

(15) Boebinger, G. S.; Jiang, H. W.; Pfeiffer, L. N.; West, K. W. Magnetic-Field-Driven Destruction of Quantum Hall States in a Double Quantum Well. *Phys. Rev. Lett.* **1990**, *64* (15), 1793–1796.

(16) Suen, Y. W.; Jo, J.; Santos, M. B.; Engel, L. W.; Hwang, S. W.; Shayegan, M. Missing Integral Quantum Hall Effect in a Wide Single Quantum Well. *Phys. Rev. B* **1991**, *44* (11), 5947–5950.

(17) Shinno, H.; Yoshizaki, R.; Tanaka, S.; Doi, T.; Kamimura, H. Conduction Band Structure of Tellurium. *J. Phys. Soc. Japan* **1973,** *35* (2), 525-533.

(18) Niu, C.; Qiu, G.; Wang, Y.; Zhang, Z.; Si, M.; Wu, W.; Ye, P. D. Gate-Tunable Strong Spin-Orbit Interaction in Two-Dimensional Tellurium Probed by Weak Antilocalization. *Phys. Rev. B* **2020**, *101* (20), 205414.

(19) Doi, T.; Nakao, K.; Kamimura, H. The Valence Band Structure of Tellurium. I. The K·p Perturbation Method. *J. Phys. Soc. Japan* **1970,** *28* (1), 36-43.

(20) Ando, T. Theory of Magnetoresistance in Tellurene and Related Mini-Gap Systems. *J. Phys. Soc. Japan* **2021,** *90* (4), 044711.

(21) Hirayama, M.; Okugawa, R.; Ishibashi, S.; Murakami, S.; Miyake, T. Weyl Node and Spin Texture in Trigonal Tellurium and Selenium. *Phys. Rev. Lett.* **2015**, *114* (20), 206401.

(22) Sakano, M.; Hirayama, M.; Takahashi, T.; Akebi, S.; Nakayama, M.; Kuroda, K.; Taguchi, K.; Yoshikawa, T.; Miyamoto, K.; Okuda, T.; et al. Radial Spin Texture in Elemental Tellurium with Chiral Crystal Structure. *Phys. Rev. Lett.* **2020**, *124* (13), 136404.

(23) Gatti, G.; Gosálbez-Martínez, D.; Tsirkin, S. S.; Fanciulli, M.; Puppin, M.; Polishchuk, S.; Moser, S.; Testa, L.; Martino, E.; Roth, S.; et al. Radial Spin Texture of the Weyl Fermions in Chiral Tellurium. *Phys. Rev. Lett.* **2020**, *125* (21), 216402.

(24) Agapito, L. A.; Kioussis, N.; Goddard, W. A.; Ong, N. P. Novel Family of Chiral-

Based Topological Insulators: Elemental Tellurium under Strain. *Phys. Rev. Lett.* **2013**, *110* (17), 176401.

(25) Tsirkin, S. S.; Puente, P. A.; Souza, I. Gyrotropic Effects in Trigonal Tellurium Studied from First Principles. *Phys. Rev. B* **2018**, *97* (3), 035158.

(26) Şahin, C.; Rou, J.; Ma, J.; Pesin, D. A. Pancharatnam-Berry Phase and Kinetic Magnetoelectric Effect in Trigonal Tellurium. *Phys. Rev. B* **2018**, *97* (20), 205206.

(27) Qiu, G.; Niu, C.; Wang, Y.; Si, M.; Zhang, Z.; Wu, W.; Ye, P. D. Quantum Hall Effect of Weyl Fermions in N-Type Semiconducting Tellurene. *Nat. Nanotechnol.* **2020**, *15* (7), 585–591.

(28) Berezovets, V.A.; Farbshtein, I.I.; Shelankov, A.L. Weak Localization under Lifted Spin-Degeneracy Conditions (Two-Dimensional Layer on a Tellurium Surface). *JETP Lett.* **1984**, *39*, 64.

(29) Akiba, K.; Kobayashi, K.; Kobayashi, T. C.; Koezuka, R.; Miyake, A.; Gouchi, J.; Uwatoko, Y.; Tokunaga, M. Magnetotransport Properties of Tellurium under Extreme Conditions. *Phys. Rev. B* **2020**, *101*, 245111.

(30) Qiu, G.; Si, M.; Wang, Y.; Lyu, X.; Wu, W.; Ye, P. D. High-Performance Few-Layer Tellurium CMOS Devices Enabled by Atomic Layer Deposited Dielectric Doping Technique. *Device Research Conference (DRC)* **2018**.

(31) Wang, Y.; Qiu, G.; Wang, R.; Huang, S.; Wang, Q.; Liu, Y.; Du, Y.; Goddard, W. A.; Kim, M. J.; Xu, X.; et al. Field-Effect Transistors Made from Solution-Grown Two-Dimensional Tellurene. *Nat. Electron.* **2018**, *1* (4), 228-236.

(32) Datta, S. Electronic Transport in Mesoscopic Systems *Cambridge university press* **1997**.

(33) Tran, S., Yang, J., Gillgren, N., Espiritu, T., Shi, Y., Watanabe, K., Taniguchi, T.,

Moon, S., Baek, H., Smirnov, D.; Bockrath, M. Surface transport and quantum Hall effect in ambipolar black phosphorus double quantum wells. *Science Advances* **2017**, *3* (6), 1603179.

(34) Muraki, K.; Saku, T.; Hirayama, Y.; Kumada, N.; Sawada, A.; Ezawa, Z. F. Interlayer Charge Transfer in Bilayer Quantum Hall States at Various Filling Factors. *Solid State Commun.* **1999**, *112* (11), 625–629.

METHODS

**Growth of 2D Te flake.** The 2D Te flakes are grown by hydrothermal method. 0.09 g of $Na_2TeO_3$ and 0.5 g of polyvinylpyrrolidone (PVP) (Sigma-Aldrich) were dissolved in 33 ml double-distilled water under magnetic stirring to form a homogeneous solution. 3.33 ml of aqueous ammonia solution (25-28%, w/w%) and 1.67 ml of hydrazine hydrate (80%, w/w%) were added to the solution. The mixture was sealed in a 50 ml Teflon-lined stainless steel autoclave and heated at 180 °C for 30 hours before naturally cooling down to room temperature.

**Device fabrication.** Te flakes were transferred onto 90 nm $SiO_2$/Si substrate. The six-terminal Hall-bar devices were patterned using two step electron beam lithography and 50 nm Ti, 20/60 nm Ti/Au metal contacts were deposited by electron beam evaporation. 20 nm ALD $Al_2O_3$ which served as top gate dielectric was deposited onto the Te flakes at 200 °C using $(CH_3)_3Al$ (TMA) and $H_2O$ as precursors. The top gate metal (50 nm Au) was deposited using electron beam evaporation after electron beam lithography patterning.

**Low temperature magneto-transport measurements.** The magneto-transport measurements were performed in a Triton 300 (Oxford Instruments) dilution fridge system with 12 T superconducting coils at temperature down to 50 mK. The high magnetic field data were collected in a 31 T resistive magnet system (Cell 9) at the National High Magnetic Field Laboratory (NHMFL) in Tallahassee, FL. The electrical data were acquired by standard small signal AC measurement technique using SR830 lock-in amplifiers (Stanford Research) at the frequency of 13.333Hz.

ASSOCIATED CONTENT

**Supporting Information**

Additional details for mobility and carrier density, comparison between $R_{xx}$ and $-\partial^2 R_{xx}/\partial B^2$, Landau level crossing at high filling factor, top gate dependence of Landau level mapping at 8 K, bilayer quantum oscillation at different magnetic field, back gate dependence of top layer Landau level mapping and top gate dependence of total Hall density at $V_{bg} = 10$ V are in the supporting information.

The authors declare no competing financial interest.

AUTHOR INFORMATION

**Corresponding Author**

*E-mail: yep@purdue.edu

**Author Contributions**

P.D.Y. conceived and supervised the project. P.D.Y. and C.N. designed the experiments. Y.W. synthesized the material under the supervision of W.W. C.N. and G.Q. fabricated the devices. C.N. and G.Q. performed the magneto-transport measurements. C.N., G.Q. and M.S. analyzed the data. P.D.Y. and C.N. wrote the manuscript and all the authors commented on it.

ACKNOWLEDGEMENTS

P.D.Y. was supported by NSF/AFOSR 2DARE program, ARO and SRC. W.W. acknowledges the School of Industrial Engineering at Purdue University for the Ravi and Eleanor Talwar Rising Star Professorship support. W.W. and P.D.Y. were also supported by NSF under grant no. CMMI-1762698. A portion of this work was performed at the National High Magnetic Field Laboratory, which is supported by National Science Foundation Cooperative Agreement No. DMR-1644779 and the State of Florida. C.N. and G.Q. acknowledge technical support from National High Magnetic Field Laboratory Staff J. Jaroszynski, A. Suslov and W. Coniglio.

FIGURES

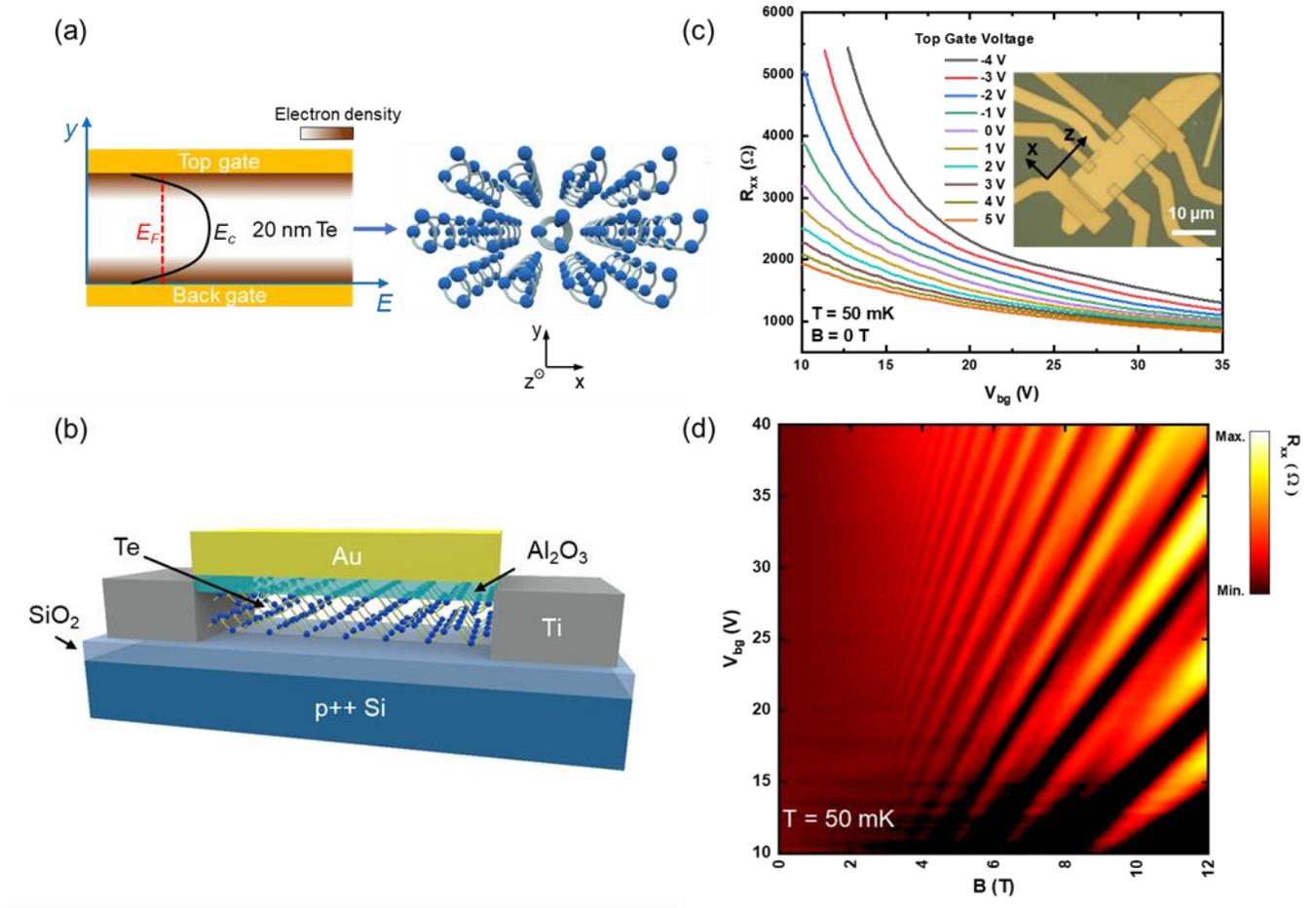

**Figure 1.** Structure and electrical transport of double-gated n-type 2D Tellurium devices. (a) Band diagram of Te bilayer electron system and the crystal structure of Te. (b) The schematic device structure of Te field-effect transistor (FET) with 20nm $Al_2O_3$ and 90 nm $SiO_2$ as gate dielectric. (c) Top and back gate voltage dependence of n-type Te channel resistance, measured at 50 mK. Inset: an optical image of double-gated Hall-bar 2D Te FET. (d) Color mapping of $R_{xx}$ by sweeping both back gate voltage $V_{bg}$ and magnetic field B at $V_{tg}$ = 0 V.

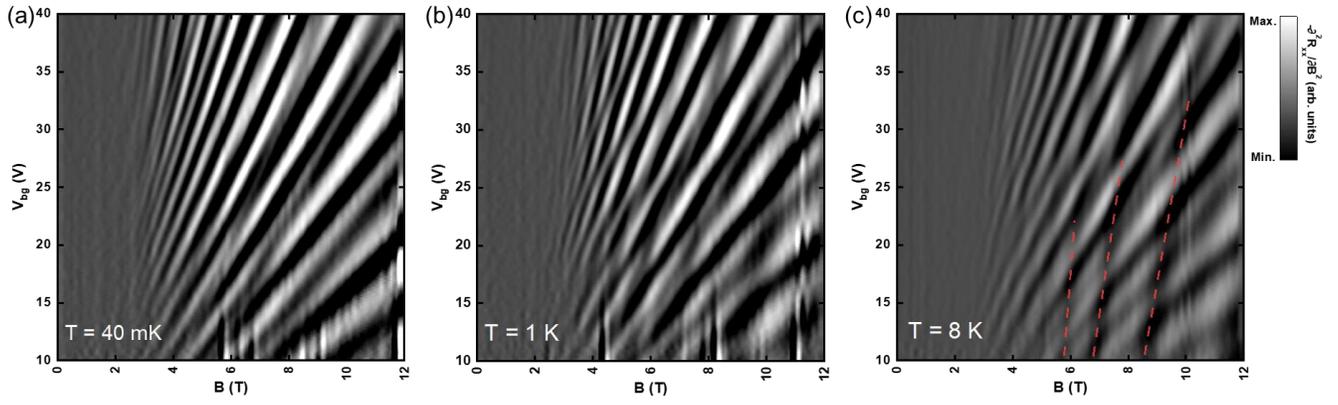

**Figure 2.** Color mapping of $-\partial^2 R_{xx}/\partial B^2$ by changing the back gate voltage $V_{bg}$ and magnetic field B under various temperatures: 40 mK (a), 1 K (b) and 8 K (c). Top gate voltage $V_{tg}$ is fixed at 0 V.

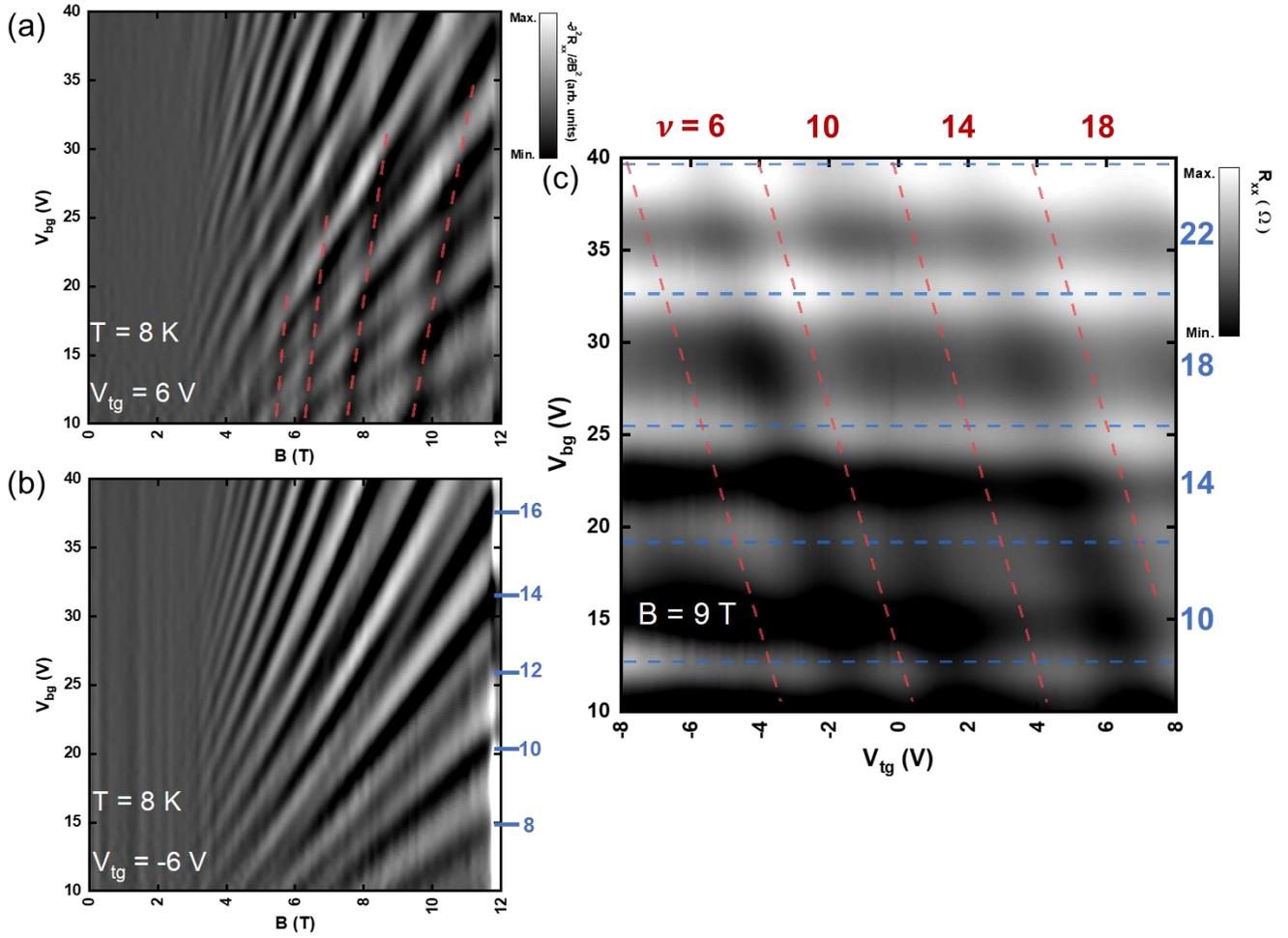

**Figure 3.** Bilayer quantum oscillations at T = 8 K. Color mapping of $-\partial^2 R_{xx}/\partial B^2$ by changing the back gate voltage $V_{bg}$ and magnetic field B at $V_{tg}$ = 6 V (a) and -6 V (b). Quantum oscillations (red dash lines) can be tuned by the top gate voltage $V_{tg}$, indicating that the oscillations originate from the top layer 2DEG. (c) Color mapping of $R_{xx}$ by sweeping both $V_{tg}$ and $V_{bg}$ at B = 9 T. Two sets of Landau levels from the top (red dash lines) and bottom (blue dash lines) 2DEG controlled independently by $V_{tg}$ and $V_{bg}$ are observed.

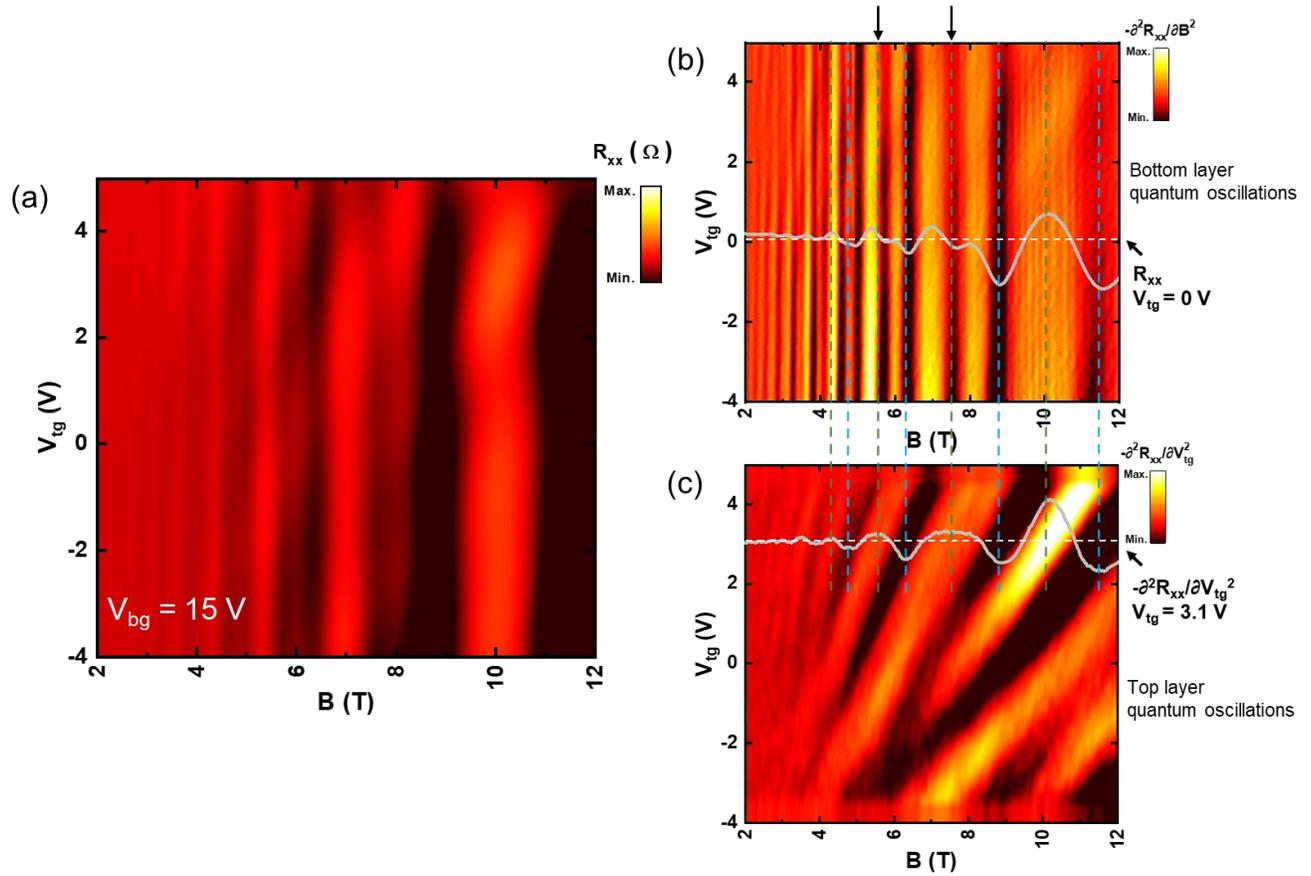

**Figure 4.** Color mapping of $R_{xx}$ (a), $-\partial^2 R_{xx}/\partial B^2$ (b), and $-\partial^2 R_{xx}/\partial V_{tg}^2$ (c) by sweeping the top gate voltage $V_{tg}$ and magnetic field B at $V_{bg} = 15$ V. The temperature is 50 mK. (b) Quantum oscillations from bottom layer 2DEG. The gray curve is $R_{xx}$ data at $V_{tg} = 0$ V. (c) Quantum oscillations from top layer 2DEG. The gray curve is $-\partial^2 R_{xx}/\partial V_{tg}^2$ data at $V_{tg} = 3.1$ V. The green and blue eye guidelines are peak and dip positions of top layer SdH oscillations at $V_{tg} = 3.1$ V.

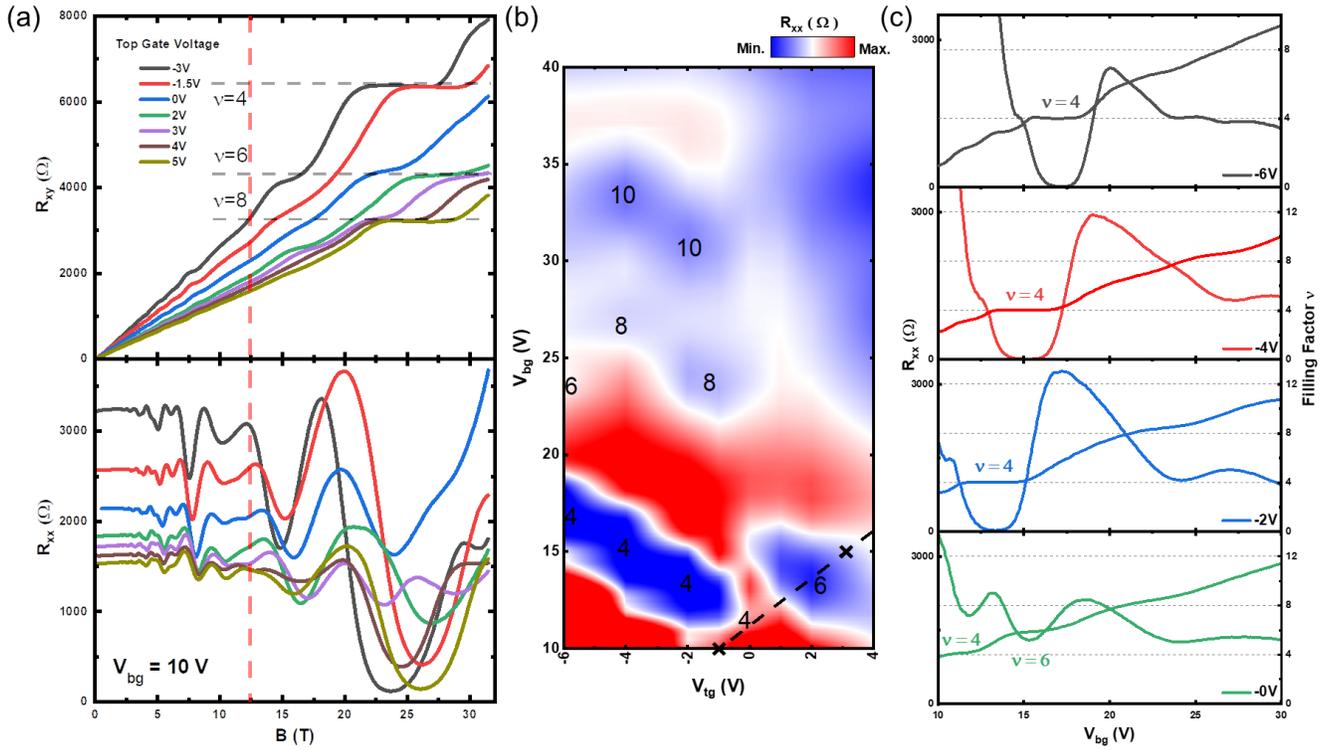

**Figure 5.** Compound and charge-transferable states in double-gated n-type 2D Te FETs. The temperature is 300 mK. (a) Transverse ($R_{xy}$ upper figure) and longitudinal ($R_{xx}$ lower figure) resistance as a function of magnetic field B. The back gate voltage is fixed at 15 V. Bilayer quantum Hall states are observed at filling factor $\nu = 4$, 6 and 8. (b) Color mapping of $R_{xx}$ by changing both top and back gate voltage at the magnetic field of 31 T. The filling factor of each quantum Hall state is indicated in the color mapping. The top and bottom charges are balanced along the dashed line. (c) Longitudinal resistance ($R_{xx}$) as a function of back gate voltage at the magnetic field of 31 T by changing the top gate voltage from -6 V to 0 V.

TOC Graphic

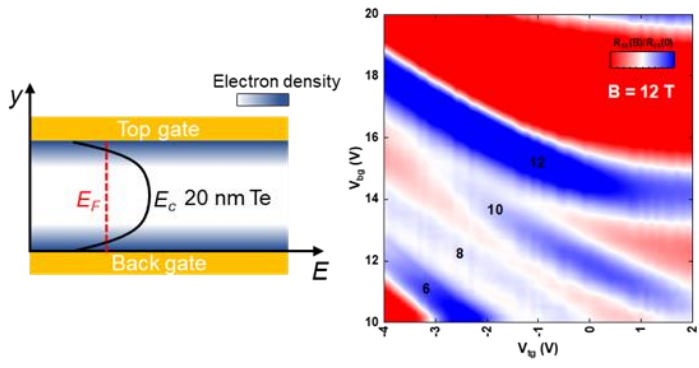

Supporting Information for:

# Bilayer Quantum Hall States in an n-type Wide Tellurium Quantum Wells

Chang Niu[1,2], Gang Qiu[1,2], Yixiu Wang[3], Mengwei Si[1,2], Wenzhuo Wu[3]

and Peide D. Ye[1,2,*]

[1]School of Electrical and Computer Engineering, Purdue University, West Lafayette, IN 47907, United States

[2]Birck Nanotechnology Center, Purdue University, West Lafayette, IN 47907, United States

[3]School of Industrial Engineering, Purdue University, West Lafayette, IN 47907, United States

* Address correspondence to: yep@purdue.edu (P.D.Y.)

## 1. Original and second derivative Landau level mapping

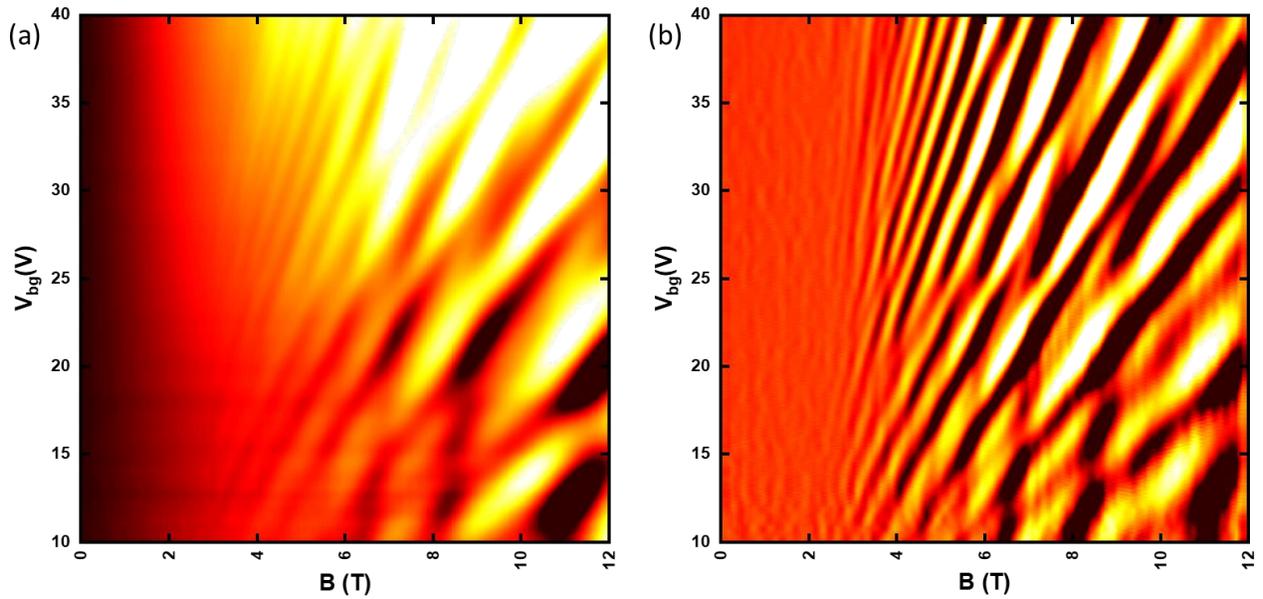

**Figure S1** Back gate voltage and magnetic field mapping of longitudinal resistance at 5K and $V_{tg}$=0V. The bright areas are resistance maxima which are crossing points formed by top and bottom layer Landau levels. (a) Original data ($R_{xx}$) mapping. (b) The second derivative of $R_{xx}$ with respect to B field ($-\partial^2 R_{xx}/\partial B^2$) mapping shows the same information with more clarity compared to original data ($R_{xx}$).

## 2. Landau level hybridization between top and bottom layer 2DEGs

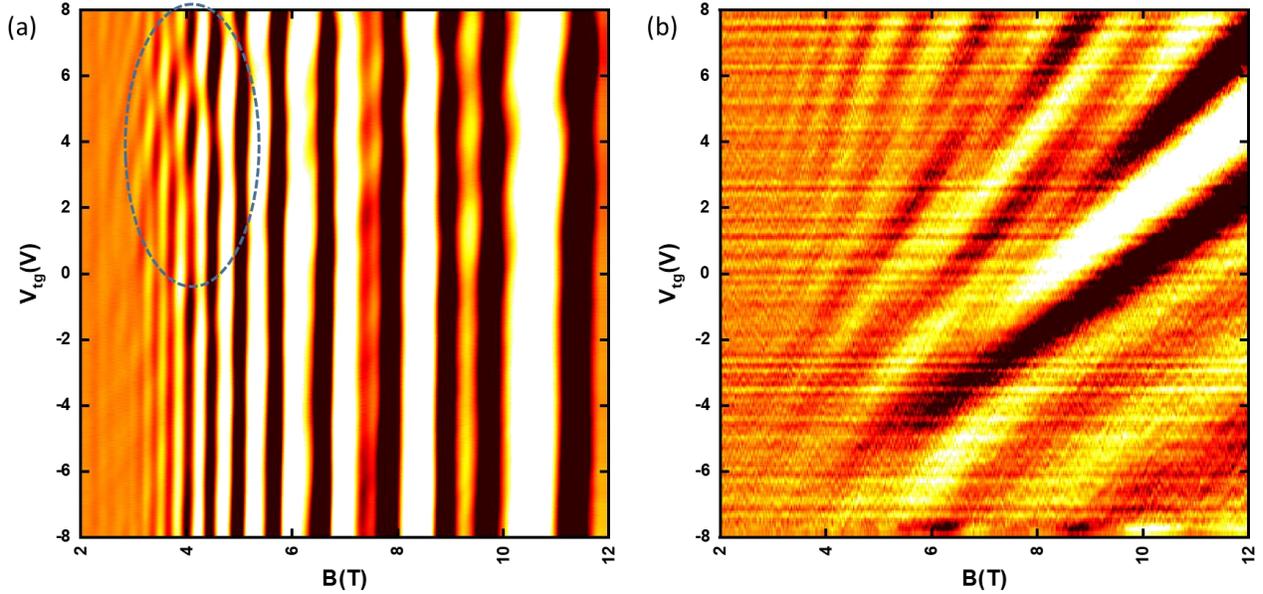

**Figure S2** (a) Top gate and magnetic field mapping of the second derivative of $R_{xx}$ with respect to magnetic field ($-\partial^2 R_{xx}/\partial B^2$). (b) Top gate and magnetic field mapping of the second derivative of $R_{xx}$ with respect to top gate voltage ($-\partial^2 R_{xx}/\partial V_{tg}^2$). The bottom layer Landau level distortion (circled in blue) can be explained by the interaction between top layer and bottom layer.

Figure S2(a) and (b) shows the quantum oscillation of Landau levels from bottom and top layer 2DEGs respectively. Landau level crossing features are clearly seen in the blue circle area. By comparing the Landau level sequences between top and bottom layer we can conclude that the crossing features are originated from the interaction between quantum oscillations.

## 3. Magnetic field B dependence of bilayer quantum oscillations

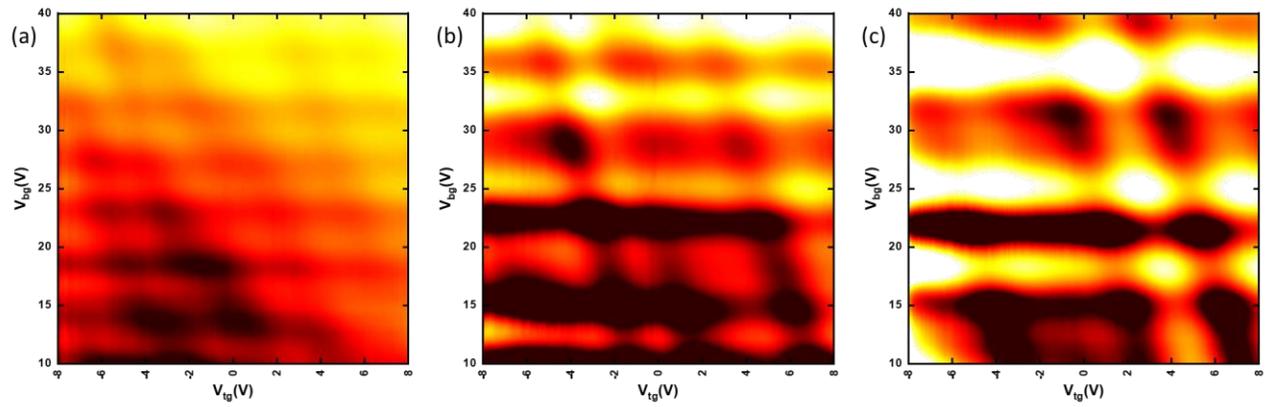

Figure S3: Top gate voltage and back gate voltage mapping of Rxx at different magnetic field: (a) 6T (b) 9T (c) 12T. The temperature is 8K.

## 4. Top gate dependence of bottom layer Landau level mapping

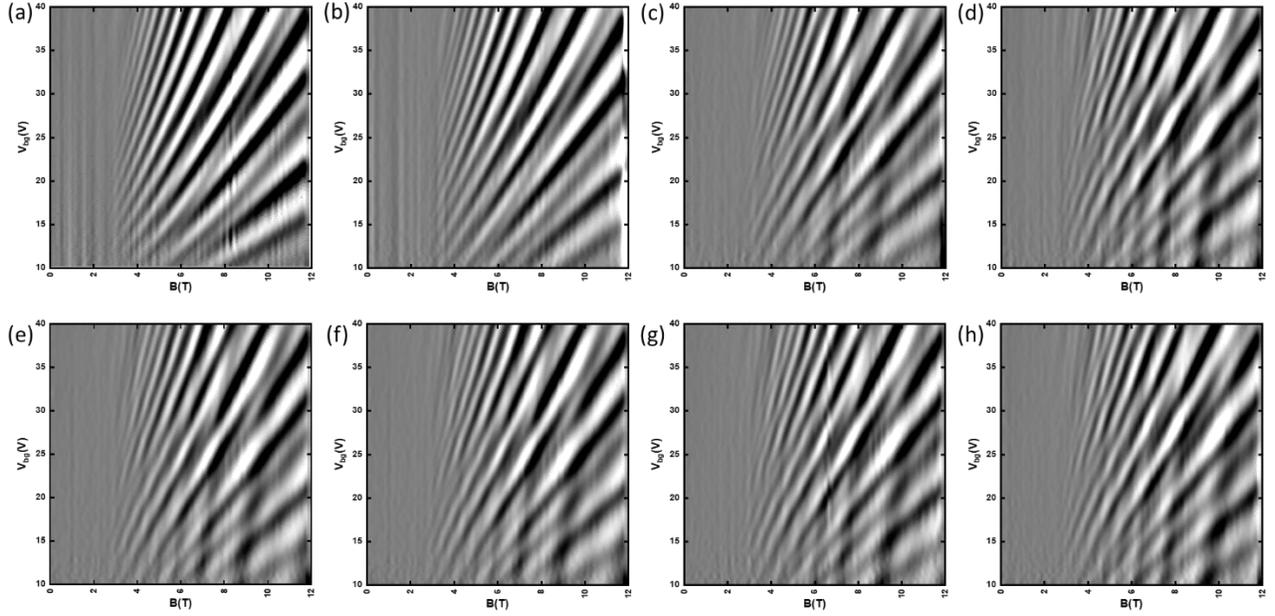

Figure S4: (a)-(h) Back gate and magnetic field mapping of the second derivative of $R_{xx}$ with respect to B field ($-\partial^2 R_{xx}/\partial B^2$) at different top gate voltage, ranging from -8V to 6V with 2V step size. The temperature is 8K.

Quantum oscillations from top layer 2DEG gradually appears with increasing oscillation frequency from $V_{tg}$ = -8 V to $V_{tg}$ = 6 V.

## 5. Back gate dependence of top layer Landau levels

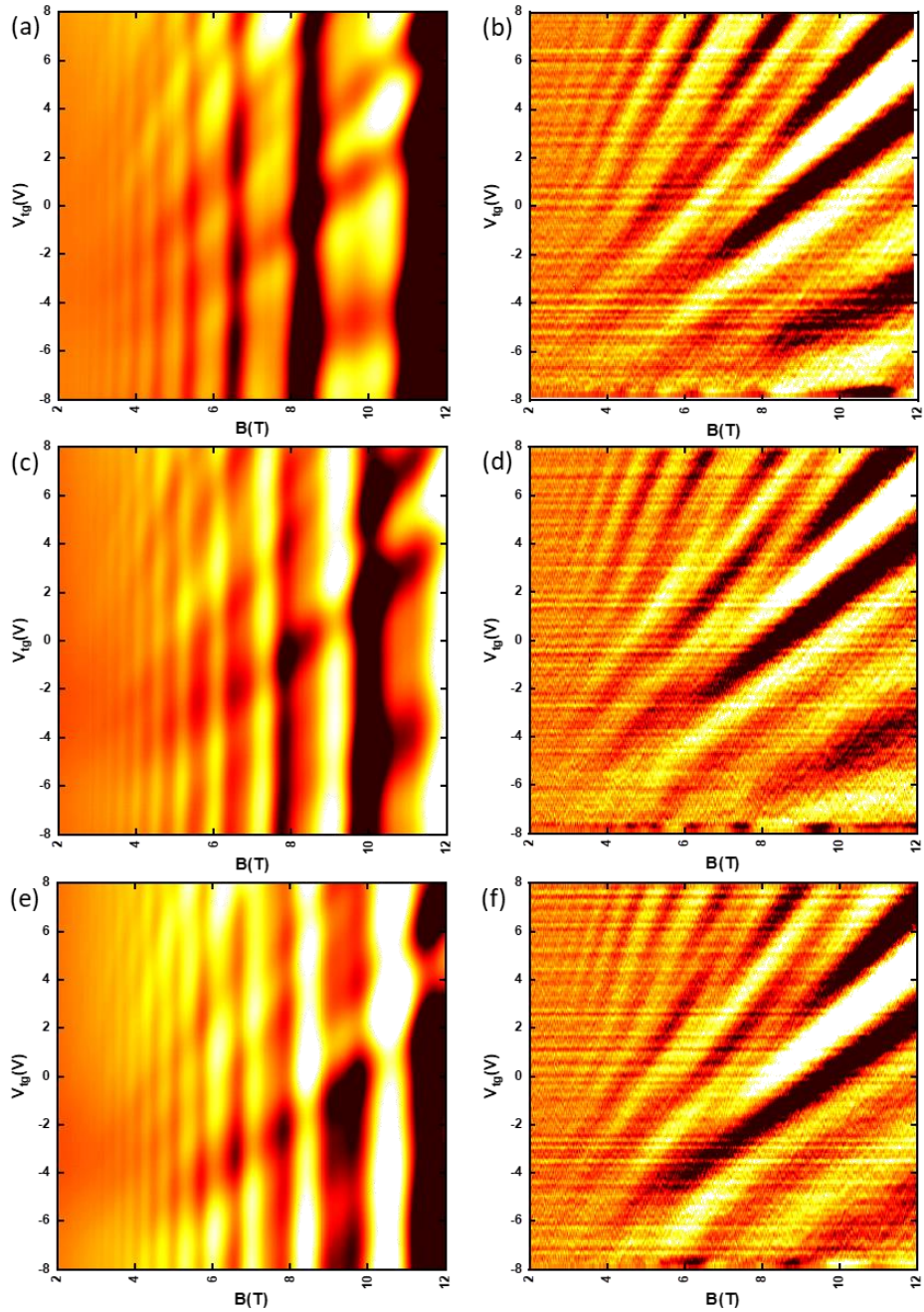

**Figure S5** Top gate voltage and magnetic field mapping of $R_{xx}$ at different back gate voltage, (a) 20V (c) 25V (d) 30V. Top gate voltage and magnetic field mapping of the second derivative of $R_{xx}$ with respect to top gate voltage ($-\partial^2 R_{xx}/\partial V_{tg}^2$) at different back gate voltage, (b) 20V (d) 25V (f) 30V. The temperature is 3K.

## 6. Total carrier density calculated from Hall measurement

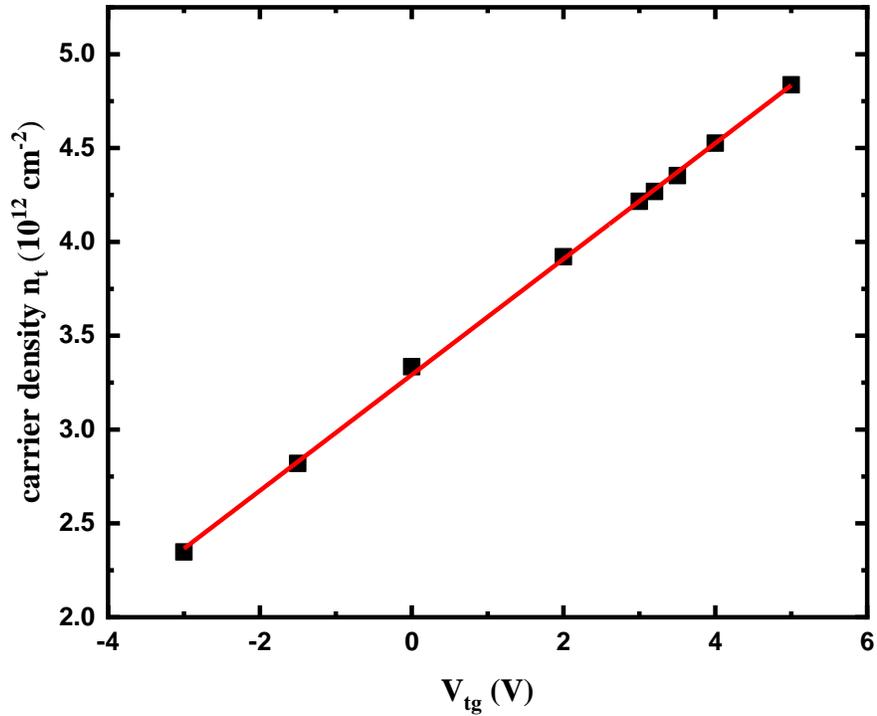

**Figure S6** Total electron density (top and bottom layer) calculated from Hall data

The carrier density of the double gated Te FET at fixed $V_{bg} = 10$ V is calculated from Hall measurement at different $V_{tg}$. We use the Hall data under 5 T to avoid the influence of quantum Hall effect.

# 7. Charge transferable states at higher filling factors

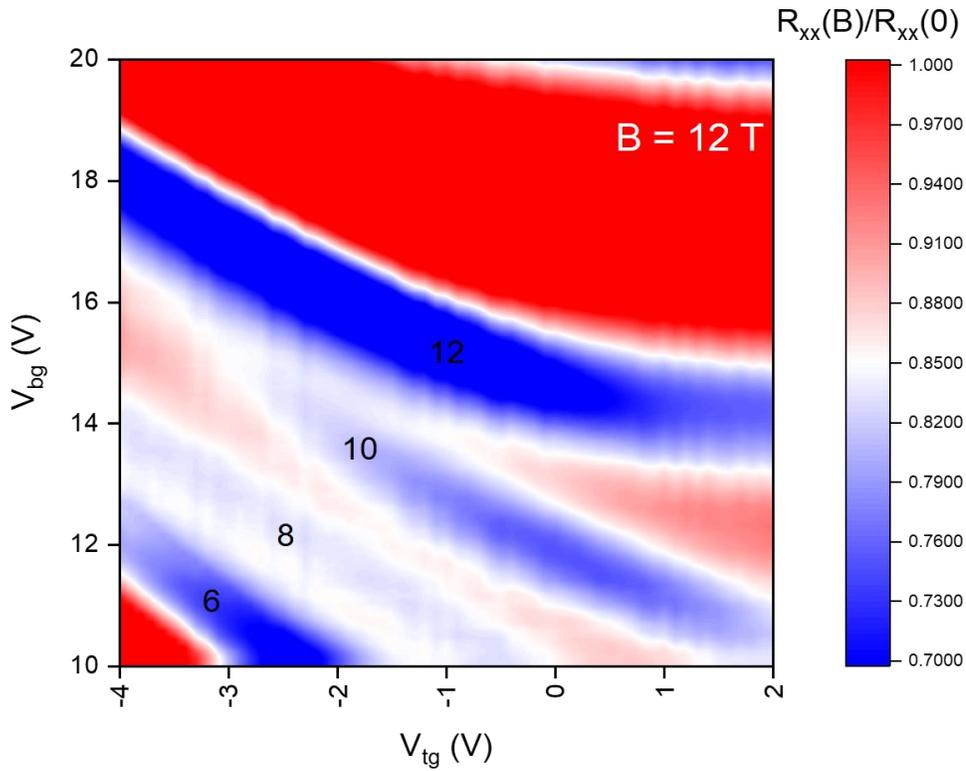

**Figure S7** Color mapping of $R_{xx}$ by changing both top and back gate voltage at the magnetic field of 12 T. The Landau levels (filling factor 6, 8, 10 and 12) are controlled by both top and back gate voltage.

Figure S7 shows the color mapping of $R_{xx}$ in another similar double-gated 2D Te device. The continues quantum Hall states of filling factor 6, 8, 10 and 12 indicate the charge transferable states in 2D Te wide quantum wells.

## 8. Topological non-trivial π Berry phase in quantum Hall sequences in both top and bottom layer electrons.

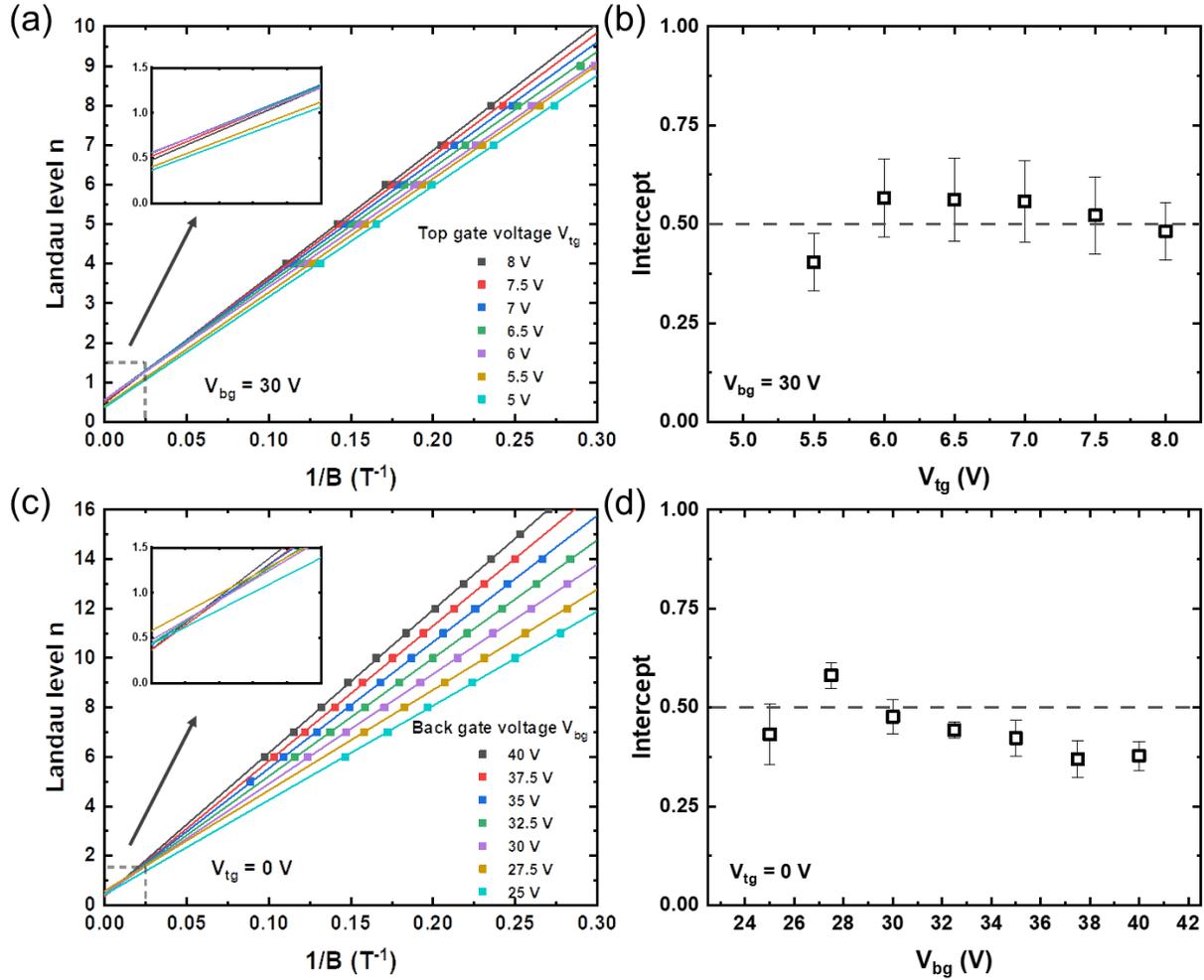

**Figure S8** (a) Landau fan diagram under different top gate voltages $V_{tg}$. The back gate voltage is fixed at 30 V. Inset: magnified view of grey dashed box in (a) with linear fittings extrapolated to the y axis. (b) Intercept of linear fitting versus top gate voltages. The grey dashed line corresponds to the π Berry phase. (c) Landau fan diagram under different back gate voltages $V_{bg}$. (d) Intercept of linear fitting versus back gate voltages. The minima in quantum oscillations are assigned to integer Landau level n.

The top layer quantum oscillations can be extracted from the mapping of the $R_{xx}$ by sweeping both the magnetic field B and the top gate voltage $V_{tg}$. The threefold screw symmetry of the Tellurium crystal protected the Weyl node at the edge of the conduction band[1,2]. Because of the Weyl node at H point, the cyclotron picks up a π Berry phase in the quantum Hall sequences.

In Figure S8, we extracted the quantum oscillations from both top and bottom layer 2DEG. We assigned the integer Landau level n to the minima of the quantum oscillations. The 0.5 intercept in Landau fan diagram (top layer Figure S8(a and b), back layer Figure S8(c and d)) indicates the π Berry phase caused by the Weyl fermions.

References

(1) Qiu, G.; Niu, C.; Wang, Y.; Si, M.; Zhang, Z.; Wu, W.; Ye, P. D. Quantum Hall Effect of Weyl Fermions in N-Type Semiconducting Tellurene. *Nat. Nanotechnol.* **2020**, *15* (7), 585–591.

(2) Hirayama, M.; Okugawa, R.; Ishibashi, S.; Murakami, S.; Miyake, T. Weyl Node and Spin Texture in Trigonal Tellurium and Selenium. *Phys. Rev. Lett.* **2015**, *114* (20), 206401.